\documentclass[twocolumn, nofootinbib]{revtex4}
\begin{document}
\title{Classical-mechanical models without observable trajectories and the Dirac electron}

\author{A. A. Deriglazov} \email{alexei.deriglazov@ufjf.edu.br}
\altaffiliation{On leave of absence from Dep. Math. Phys., Tomsk Polytechnical University, Tomsk, Russia.}

\affiliation{Depto. de Matem\'atica, ICE, Universidade Federal de Juiz de Fora, MG, Brazil}

\begin{abstract}
We construct a non-Grassmann spinning-particle model which, by analogy with quantum mechanics, does not admit the
notion of a trajectory within the position space. The pseudo-classical character of the model allows us to avoid the
inconsistencies arising in the quantum-mechanical interpretation of a one-particle sector of the Dirac equation.
\end{abstract}

\maketitle

\section{Introduction}
Non-abelian gauge groups play a crucial role in field theory as well as in the Standard model. In this work we
observe that they imply new possibilities when used in the construction of finite-dimensional theories as well. We
suggest and discuss the notion of pseudo-classical mechanics (pCM), the term by which we refer to models with a
number of observable configuration-space variables less than the number of physical degrees of freedom. In other words,
we consider classical-mechanical models which, by analogy with quantum mechanics, do not admit the notion of a
trajectory within the position (i.e. configuration) space. So we expect that pCM turns out to be useful in describing quantum phenomena by (semi) classical methods [1-11]. Classical mechanics with such a strange property can be
constructed on the basis of a singular Lagrangian with a multi-parametric group of local symmetries. As examples of
pCM we present the models which are invariant under transformation of the non-abelian gauge group with two and three local parameters. Symmetries
imply functional ambiguity in solutions to equations of motion: besides the integration constants $c_i$, the solution
depends on the arbitrary functions $e_a(\tau)$, $x=f(\tau, c_i, e_a(\tau))$. According to the general theory of singular
systems [12-14], variables with ambiguous dynamics do not represent observable quantities. So, when we are
dealing with the locally-invariant theory, our first task is to find candidates for observables, which are variables
with unambiguous dynamics. Equivalently, we can look for the gauge-invariant variables.

We start in Section 2 with a couple of toy models and show that, generally, it is impossible to construct the
observables within the position variables only. It is worth noting that on the phase space there always is the
well-defined notion of a trajectory [13]. In Section 3 we consider a more realistic case, presenting the non-Grassmann
model of the Dirac electron. In Section 4 we show how the pseudo-classical character of the model allows us to solve
the problems arising [2, 3, 18] when we try to apply the methods of relativistic quantum mechanics to a one-particle
sector of the Dirac equation.

\section{Toy models}
One of the local symmetries which will be presented in our models is reparametrization invariance. So, we first
outline the reparametrization invariant formulation of a relativistic particle.

The motion of a particle in special-relativity theory can be described starting from the three-dimensional action
$-mc\int\ dt\sqrt{c^2-(\frac{dx^i}{dt})^2}$. The problem here is that the Lorentz transformations,
$x'^{\mu}=\Lambda^\mu{}_\nu x^\nu$, act on the physical dynamical variables $x^i(t)$ in a higher nonlinear way. To
improve this, we pass from three-dimensional to four-dimensional formulation. Introducing the parametric
representation $x^\mu(\tau)=(ct(\tau), x^i(\tau))$ for the trajectory $x^i(t)$, the particle can be described by the
Lagrangian action
\begin{eqnarray}\label{2.2}
S=\int d\tau(\frac{1}{2 e}(\dot x^\mu)^2-\frac{e}{2}m^2c^2).
\end{eqnarray}
The corresponding Hamiltonian action reads
\begin{eqnarray}\label{2.3}
S_H=\int d\tau p_\mu\dot x^\mu+p_e\dot e-\frac12e(p^2+m^2c^2)-\lambda_e p_e,
\end{eqnarray}
where $\lambda_e(\tau)$ stands for the Lagrangian multiplier of the primary constraint $\pi_e=0$. Variation of the
functional implies the Hamiltonian equations
\begin{eqnarray}\label{2.4}
\dot e=\lambda_e, \qquad \dot p_e=0, \qquad \dot x^{\mu}=ep^{\mu}, \qquad \dot p_{\mu}=0,
\end{eqnarray}
as well as the constraints $\pi_e=0$, $p^2+m^2c^2=0$. We note that the variable $\lambda_e(\tau)$ cannot be determined
with the constraints, nor with the dynamical equations. As a consequence (see the first of Eqs. (\ref{2.4})),
the variable $e$ turns out to be an arbitrary function as well. Since $e(\tau)$ enters into the equation for $x^{\mu}$,
its general solution contains, besides the arbitrary integration constants, the arbitrary function $e(\tau)$. Hence the
only unambiguous ones among the initial variables are $p^\mu$ and $\pi_e$, see Eqs. (\ref{2.4}). $x^\mu$ has one-parameter
ambiguity due to $e$.

The ambiguity reflects the freedom in the choice of parametrization for the particle trajectory ($\alpha(\tau)$ is an
infinitesimal  function)
\begin{equation}\label{2.5}
\begin{array}{c}
\tau\rightarrow\tau'=\tau-\alpha, \qquad \qquad \\
x^\mu(\tau)\rightarrow x'{}^\mu(\tau')=x^\mu(\tau), \quad \\
e(\tau)\rightarrow e'(\tau')=(1+\dot\alpha)e(\tau).
\end{array}
\mbox{then} ~
\begin{array}{c}
\delta x^{\mu}=\alpha\dot x^{\mu}, \\
\quad\delta e=(\alpha e)\dot{}.
\end{array}
\end{equation}
Action (\ref{2.2}) turns out to be invariant under the reparametrizations.

By construction, the expression for the physical trajectory
$x^i(t)$ is obtained resolving the equation $x^0=x^0(\tau)$ with
respect to $\tau$, $\tau=\tau(x^0)$, then $x^i(t)\equiv
x^i(\tau(x^0))$. Using the expression
\begin{eqnarray}\label{2.6.1}
\frac{df}{dx^0}=\frac{\dot f(\tau)}{\dot x^0(\tau)},
\end{eqnarray}
for the derivative of a function given in parametric form, we
obtain
\begin{eqnarray}\label{2.6}
\frac{dx^i}{dt}= c\frac{\dot x^i}{\dot x^0} =
c\frac{p^i}{\sqrt{\vec p^2+m^2c^2}},
\end{eqnarray}
Eq. (\ref{2.6}) coincides with that of the three-dimensional formulation. As should be the case, the physical coordinate $x^i(t)$
has unambiguous evolution. \par \noindent {\bf Toy model which admits the position-space trajectories.} Consider the
Lagrangian action
\begin{eqnarray}\label{2.7}
S=\int d\tau\frac{1}{2e_1}(\dot x^\mu-e_2x^\mu)^2.
\end{eqnarray}
This is written on the configuration space $x^\mu$, $e_1$ and $e_2$; the Minkowski metric is $\eta^{\mu\nu}=(-, +, +,
+)$. The action is invariant under the reparametrizations, $\delta x^\mu=\alpha\dot x^\mu$, $\delta e_1=(\alpha
e_1)\dot{}$, $\delta e_2=(\alpha e_2)\dot{}$, as well as under the following transformations with the parameter
$\beta(\tau)$
\begin{eqnarray}\label{2.8}
\delta x^\mu=\beta x^\mu, \qquad \delta e_1=2\beta e_1, \qquad \delta e_2=\dot\beta.
\end{eqnarray}
The transformations form a non-abelian group, $[\delta_\alpha, \delta_\beta]=\delta_{\tilde\beta}$,
$\tilde\beta=-\alpha\dot\beta$. Each local symmetry removes two degrees of freedom \cite{aad5}, so the number of configuration-space observables is equal to 2.

The Hamiltonian of the theory (\ref{2.7}) is
\begin{eqnarray}\label{2.9}
H=\frac{e_1}{2}p^2+e_2(px).
\end{eqnarray}
This implies the Hamiltonian equations (in what follows, we omit equations for the auxiliary variables $e_k$, $p_{ek}$,
as they are not necessary for discussion of the ambiguity of the variable $x^\mu$ )
\begin{eqnarray}\label{2.10}
\dot x^\mu=e_1p^\mu+e_2 x^\mu, \qquad \dot p^\mu=-e_2p^\mu,
\end{eqnarray}
as well as the constraints
\begin{eqnarray}\label{2.11}
p^2=0, \qquad (px)=0.
\end{eqnarray}
The equation for $x^\mu$ has two-parametric ambiguity due to $e_1$ and $e_2$, while that for $p^\mu$ has one-parametric
ambiguity. Inside the light-cone, we construct the variables
\begin{eqnarray}\label{2.12}
\tilde x^\mu=\frac{x^\mu}{\sqrt{-x^2}}, \qquad \tilde p^\mu=\sqrt{-x^2}p^\mu.
\end{eqnarray}
Their equations read
\begin{eqnarray}\label{2.13}
\dot{\tilde x}^\mu=e_1\frac{p^\mu}{\sqrt{-x^2}}, \qquad \dot{\tilde p}^\mu=0.
\end{eqnarray}
They can be compared with Eqs. (\ref{2.4}). The constraints
(\ref{2.11}) acquire the form $\tilde p^2=0$, $(\tilde p \tilde
x)=0$. Note that $\tilde x$ is $\beta$\,-invariant variable. So
the ambiguity presented in Eq. (\ref{2.13}) is due to the
reparametrization symmetry. In accordance with this observation,
we assume that the functions $\tilde x^\mu(\tau)$ and $\tilde
p^\mu(\tau)$ represent the reparametrization-invariant variables
$\tilde x^i(t)$ and $\tilde p^\mu(t)$ in the parametric form.
Their equations of motion read
\begin{eqnarray}\label{2.14}
\frac{d\tilde x^i}{dt}=c\frac{p^i}{p^0}\equiv c\frac{\tilde p^i}{\tilde p^0}, \qquad \frac{d\tilde p^\mu}{dt}=0.
\end{eqnarray}
Since they are unambiguous, the variables $\tilde x^i(t)$ and $\tilde p^\mu(t)$ are candidates for the observables.

By construction, $\tilde x^\mu$ obey the identity $\tilde x^\mu\tilde x_\mu=-1$. So only three of them can be taken as
coordinates of the configuration space. Adding the variable $\sigma=\frac{1}{\sqrt{-x^2}}$ to the set $\tilde x^i$, we
obtain a coordinate system. As the two independent observables we can take the gauge-invariant variables $\tilde
x^1(t)$ and $\tilde x^2(t)$. Hence, the present model admits observable trajectories constructed within the
position space; see Eq. (\ref{2.12}).
%
%

To conclude, we point out that Poisson brackets of the Lorentz-covariant observables (\ref{2.12}) generate the
non-commutative algebra
\begin{eqnarray}\label{2.15}
\{\tilde x^\mu, \tilde p^\nu\}=N^{\mu\nu}(\tilde x), ~ \{\tilde
x^\mu, \tilde x^\nu\}=0, ~ \{\tilde p^\mu, \tilde p^\nu\}=\tilde
p^{[\mu}\tilde x^{\nu]}.
\end{eqnarray}
Here and below we denote
\begin{eqnarray}\label{2.15.1}
N^{\mu\nu}(a)\equiv\eta^{\mu\nu}-\frac{a^\mu a^\nu}{a^2}.
\end{eqnarray}

\par
\noindent{\bf Toy model without position-space trajectories.} Consider the following Lagrangian action written for the
variables $x^\mu$, $\omega^\mu$, $e_1$ and $e_2$
\begin{eqnarray}\label{2.16}
S=\int d\tau\frac{1}{2(e_1-e_2^2)}\left[(Dx)^2+2e_2(Dx \dot\omega)+e_1\dot \omega^2\right]- \cr
\frac{e_1}{2}m^2c^2+\frac12\omega^2. \qquad \qquad \qquad \qquad
\end{eqnarray}
We have denoted $Dx^\mu\equiv\dot x^\mu-e_2\omega^\mu$. The
non-abelian gauge group is composed by reparametrizations as well
as by the following transformations with the parameter
$\beta(\tau)$:
\begin{eqnarray}\label{2.18}
\delta x^\mu=\frac{\beta}{e_1-e_2^2}(Dx^\mu+e_2\dot\omega^\mu), \qquad \delta e_1=\dot\beta.
\end{eqnarray}
This implies that the number of physical degrees of freedom on configuration (phase) space is equal to 6 (12).

Denoting conjugate momenta of $x$, $\omega$ by $p$, $\pi$, the Hamiltonian of the theory (\ref{2.16}) reads
\begin{eqnarray}\label{2.19}
H=\frac12\pi^2-\frac12\omega^2+\frac12e_1(p^2+m^2c^2)+e_2p_\mu(\omega^\mu-\pi^\mu).
\end{eqnarray}
This implies the Hamiltonian equations
\begin{eqnarray}\label{2.20}
\dot x^\mu=e_1p^\mu+e_2(\omega^\mu-\pi^\mu), \quad \dot p^\mu=0, \cr \dot\omega^\mu=\pi^\mu-e_2p^\mu, \quad
\dot\pi^\mu=\omega^\mu-e_2p^\mu.
\end{eqnarray}
as well as the first-class constraints
\begin{eqnarray}\label{2.21}
p^2+m^2c^2=0, \qquad (p, \omega-\pi)=0.
\end{eqnarray}
The equation for $x$ has two-parametric ambiguity due to $e_1$ and
$e_2$, while those for $\omega$ and $\pi$ have one-parametric
ambiguity.

Taking into account the first-class constraints, we could expect 6 observable dynamical variables on the configuration
space. However, it is easy to see that any configuration-space quantity $a^\mu(x, \omega)$ with one-parametric
ambiguity is proportional to $\omega^\mu$. Similarly to the previous model, this can be used to construct only three
unambiguous dynamic variables. As the six-dimensional configuration space can not be spanned with the unambiguous
variables, the model represents an example of pseudo-classical mechanics.

On the phase space we can construct various variables with
one-parametric ambiguity due to $e_2$
\begin{eqnarray}\label{2.22}
\tilde x^\mu=x^\mu-\frac{(px)}{p^2}p^\mu, \qquad \dot{\tilde
x}^\mu=e_2(\omega^\mu-\pi^\mu),
\end{eqnarray}
\begin{eqnarray}\label{2.23}
\tilde p^\mu=p^\mu, \qquad \dot{\tilde p}^\mu=0,
\end{eqnarray}
\begin{eqnarray}\label{2.24}
\tilde \omega^\mu=\frac{(\omega+\pi,
p)}{2p^2}(\omega^\mu-\pi^\mu), \quad \dot{\tilde
\omega}^\mu=-e_2(\omega^\mu-\pi^\mu),
\end{eqnarray}
\begin{eqnarray}\label{2.25}
\tilde\pi^\mu=\frac{\omega^\mu+\pi^\mu}{\sqrt{(\omega+\pi)^2}}, \qquad \dot{\tilde\pi}^\mu=
-\frac{2e_2N^{\mu\nu}(\omega+\pi)p_\nu}{\sqrt{(\omega+\pi)^2}}.
\end{eqnarray}
\begin{eqnarray}\label{2.26}
J^{\mu\nu}=\omega^\mu\pi^\nu-\omega^\nu\pi^\mu, \qquad \dot
J^{\mu\nu}=e_2p^{[\mu}(\omega-\pi)^{\nu]}.
\end{eqnarray}
The constraints (\ref{2.21}) acquire the form $\tilde p^2+m^2c^2=0$, $(\tilde p \tilde\omega)=0$. The new variables
are invariants of the $\beta$\,-transformation\footnote{Due to the identities $\tilde x^\mu\tilde p_\mu=0$, $\tilde
\pi^\mu\tilde \pi_\mu=1$ and $\epsilon^{ijk}J_{ij}J_{0k}=0$, not all of them are independent.}. So the ambiguity
presented in Eqs. (\ref{2.22})-(\ref{2.26}) is due to the reparametrization symmetry. Similarly to the case of the
relativistic particle, we assume that the functions $\tilde x^\mu(\tau)$, $\tilde p^\mu(\tau)$, $\tilde
\omega^\mu(\tau)$, $\tilde \pi^\mu(\tau)$ and $J^{\mu\nu}(\tau)$ represent the physical variables $\tilde x^i(t)$,
$\tilde p^\mu(t), \ldots$ in the parametric form. According to Eq. (\ref{2.6.1}), the dynamics of the physical variables is
unambiguous.

The Poisson-bracket algebra of the Lorentz-covariant observables is highly noncommutative; the nonvanishing brackets are
\begin{eqnarray}\label{2.27}
\{\tilde x^\mu, \tilde x^\nu\}=\frac{p^{[\mu}\tilde x^{\nu]}}{p^2}, \qquad \{\tilde x^\mu, p^\nu\}=N^{\mu\nu}(p), \cr
\{\tilde\omega^\mu, \tilde\omega^\nu\}=\frac{p^{[\mu}\tilde\omega^{\nu]}}{p^2}, \quad \{\tilde\omega^\mu,
\tilde\pi^\nu\}=\frac{(p\tilde\pi)}{p^2}N^{\mu\nu}(\tilde\pi), \cr \{\tilde x^\mu,
\tilde\omega^\nu\}=\frac{\tilde\pi^\mu\tilde\omega^\nu}{(p\tilde\pi)}-\frac{p^\mu\tilde\omega^\nu}{p^2}, \qquad \qquad
\cr \{J^{\mu\nu}, J^{\alpha\beta}\}=\eta^{\mu\alpha}J^{\nu\beta}-\eta^{\mu\beta}J^{\nu\alpha}  \qquad \qquad \cr
-\eta^{\nu\alpha}J^{\mu\beta}+\eta^{\nu\beta}J^{\mu\alpha}, \qquad \qquad \cr \{J^{\mu\nu},
\tilde\omega^\alpha\}=\eta^{\alpha[\mu}\tilde\omega^{\nu]}+\frac{p^{[\mu}\tilde\pi^{\nu]}\tilde\omega^\alpha}{(p\tilde\pi)},
\qquad  \cr \{J^{\mu\nu}, \tilde\pi^\alpha\}=\eta^{\alpha[\mu}\tilde\pi^{\nu]}. \qquad \qquad \quad
\end{eqnarray}

The set of 12 independent observables of the phase-space can be selected as follows. We parameterize the initial space
by the coordinates $x^0$, $\tilde x^i$, $\tilde p^\mu$, $\tilde\omega^\mu$, $S^i\equiv\epsilon^{ijk}J_{jk}$ and
$\gamma=\sqrt{(\omega+\pi)^2}$. The dynamics of the theory is restricted on the surface $\tilde p^2+m^2=0$, $(\tilde p
\tilde\omega)=0$ which is invariant under the action of the gauge group\footnote{We use the phase-space form of
reparametrizations, $\delta\tilde\omega^\mu=-\alpha e_2(\omega^\mu-\pi^\mu)$, $\delta p^\mu=0$; see [15] for details.}. The surface can be parameterized by $x^0$, $\tilde x^i$, $\tilde p^i$, $\tilde\omega^i$, $S^i$ and $\gamma$.
%
%
The corresponding dynamic variables $\tilde x^i(t)$, $\tilde p^i(t)$, $\tilde\omega^i(t)$ and $S^i(t)$ have
unambiguous dynamics. Hence we can take them as the independent observables.

\section{Non-Grassmann Mechanical model of the Dirac electron}
As a more realistic example of pCM, we discuss the spinning-particle model suggested in a recent work [15]. The
configuration space of the model consist of the dynamical variables $Q^\alpha(\tau)=(x^\mu, \omega^\nu, \omega^5)$ as
well as the auxiliary variables $e_l$, $l=1, 2, 3, 4$. $x^\mu$ are coordinates of the Minkowski space with the metric
$\eta^{\mu\nu}=(-, +, +, +)$. The spin-space $\omega^A=(\omega^\mu, \omega^5)$ is equipped with $SO(2, 3)$\,- metric
$\eta^{AB}=(-, +, +, +,-)$. Consider the Poincare-invariant Lagrangian
\begin{eqnarray}\label{3.1}
L=\frac12G_{\alpha\beta}\dot Q^\alpha\dot Q^\beta-\frac{e_4}{2}\omega^A\omega_A-\frac{e_l}{2}a_l.
\end{eqnarray}
We have denoted $a_1=m^2c^2, a_2=mc\hbar$, and $a_3$, $a_4$ are real numbers. In what follows, we discuss the free
theory. Interaction with an external electromagnetic field will be discussed at the end of Section 4. The kinetic term
looks like that of a free particle moving on the curved nine-dimensional space with the metric
\begin{eqnarray}\label{3.2}
G_{\alpha\beta}=\left(
\begin{array}{cccccc}
e_3G_{\mu\nu} & -e_2\omega^5G_{\mu\nu} &
\frac{e_2}{A}\omega_\mu \\
-e_2\omega^5G_{\mu\nu} & e_1G_{\mu\nu}
+\frac{e_2^2\omega_\mu\omega_\nu}{e_3A}
& -\frac{e_2^2\omega^5}{e_3A}\omega_\mu \\
\frac{e_2}{A}\omega_\nu & -\frac{e_2^2\omega^5}{e_3A}\omega_\nu & -\frac{B}{e_3A}
\end{array} \right)
\end{eqnarray}
We have denoted
$G_{\mu\nu}=\frac{1}{B}[\eta_{\mu\nu}-\frac{e_2^2\omega_\mu\omega_\nu}{A}]$,
$B=e_1e_3-e_2^2(\omega^5)^2$ and $A=B+e_2^2(\omega^\mu)^2$.

We introduce the abbreviation
\begin{eqnarray}\label{3.3}
Dx^\mu\equiv\dot
x^\mu-\frac{e_2}{e_3}(\omega^5\dot\omega^\mu-\omega^\mu\dot\omega^5),
\end{eqnarray}
then the Lagrangian (\ref{3.1}) can be written as follows,
\begin{eqnarray}\label{3.4}
L=\frac{e_3}{2}G_{\mu\nu}Dx^\mu Dx^\nu+ \frac{1}{2e_3}\dot\omega^A\dot\omega_A-
\frac{e_4}{2}\omega^A\omega_A-\frac{e_l}{2}a_l.
\end{eqnarray}
The Lagrangian is invariant under a three-parametric group of local symmetries. One of them is the reparametrization
symmetry.
%
%
Besides, there are two more symmetries with the local parameters
$\beta(\tau)$, $\gamma(\tau)$
\begin{eqnarray}\label{3.5}
\delta_\beta x^\mu=\beta p^\mu, \quad \delta_\beta e_1=\dot\beta;
\end{eqnarray}
\begin{eqnarray}\label{3.6}
\delta_\gamma\omega^A=\gamma e_3\pi^A, \quad \delta_\gamma\pi^A=-\gamma e_4\omega^A, \cr \delta_\gamma e_3=(\gamma
e_3)\dot{}, \quad \delta_\gamma e_4=(\gamma e_4)\dot{}. \qquad
\end{eqnarray}
Here $p_\mu=\frac{\partial L}{\partial\dot x^\mu}$, $\pi_A=\frac{\partial L}{\partial\dot\omega^A}$. In the Hamiltonian
formulation, the Lagrangian (\ref{3.4}) leads to the following Hamiltonian [15]
\begin{eqnarray}\label{3.7}
H=\frac{e_1}{2}(p^2+m^2c^2)+\frac{e_2}{2}(p_\mu J^{5\mu}+mc\hbar)+
\cr
\frac{e_3}{2}(\pi^A\pi_A+a_3)+\frac{e_4}{2}(\omega^A\omega_A+a_4)+\lambda_{ea}\pi_{ea},
\end{eqnarray}
where
\begin{eqnarray}\label{3.8}
J^{5\mu}=2(\omega^5\pi^\mu-\omega^\mu\pi^5).
\end{eqnarray}
If we omit the spin-space coordinates, $\omega^A=\pi_A=0$, the Hamiltonian reduces to that of the spinless particle,
see (\ref{2.3}).

The Hamiltonian implies the constraints
\begin{eqnarray}\label{3.9}
\omega^A\omega_A+a_4=0, \qquad \pi^A\omega_A=0;
\end{eqnarray}
\begin{eqnarray}\label{3.10}
p^2+m^2c^2=0, \qquad \pi^A\pi_A+a_3=0;
\end{eqnarray}
\begin{eqnarray}\label{3.11}
p_\mu J^{5\mu}+mc\hbar=0.
\end{eqnarray}
The first one states that configuration space of spin is anti-de Sitter space. The constraints (\ref{3.9}) form the
second-class pair while those of Eqs. (\ref{3.10}) and (\ref{3.11}) are the first-class constraints. The constraint
(\ref{3.11}), being imposed on the state vector, leads to the Dirac equation\footnote{The present model implies both
the Dirac equation and the mass-shell condition $p^2+m^2c^2=0$. The model without the mass-shell condition has been
discussed in [16, 17]. This shows the same undesirable properties as those of the Dirac equation in the classical limit [2,
3].}, $(\gamma^\mu\hat p_\mu+mc)\Psi=0$ (see [15] for details).

The Hamiltonian equations of the theory read
\begin{eqnarray}\label{3.12}
\dot x^\mu=e_1p^\mu+\frac12e_2J^{5\mu}, \qquad \dot p^\mu=0;
\qquad \qquad \qquad ~ \cr
\dot\omega^\mu=e_3\pi^\mu+e_2\omega^5p^\mu, \qquad
\dot\pi^\mu=e_2\pi^5p^\mu-\frac{a_3}{a_4}e_3\omega^\mu; \cr
\dot\omega^5=e_3\pi^5+e_2(p\omega), \qquad ~
\dot\pi^5=e_2(p\pi)-\frac{a_3}{a_4}e_3\omega^5. ~
\end{eqnarray}
The only unambiguous variable is $p^\mu$. The configuration space variables $x^\mu$, $\omega^A$ have two-parametric
ambiguity.

Let us compute the total number of physical degrees of freedom. Omitting the auxiliary variables and the corresponding
constraints, we have $18$ phase-space variables $x^\mu$, $p_\mu$, $\omega^A$, $\pi_A$ subject to the constraints
(\ref{3.9})- (\ref{3.11}). Taking into account that each second-class constraint rules out one variable, whereas each
first-class constraint rules out two variables, the number of physical degrees of freedom on the phase space is
$18-(2+2\times 3)=10$. Hence we could expect five observables on the configuration space. However, using the
configuration-space variables only, it is impossible to construct five unambiguous quantities. Thus, once again we have
an example of pseudo-classical mechanics.

Let us discuss the physical sector of the phase space. A brief inspection of the equations of motion allows us to construct
the Lorentz-covariant variables with one-parameter ambiguity. They are the five-dimensional angular-momentum
tensor\footnote{Note that the constraints (\ref{3.9}) and (\ref{3.10}) fix the value of the Casimir operators of the $SO(2, 3)$
group. Besides, they guarantee that $J^{5\mu}$ is the time-like vector $(J^{5
\mu})^2=-4(a_3(\omega^5)^2+a_4(\pi^5)^2)<0$, for positive values of $a_3$, $a_4$; see [15] for details.}
$J^{AB}=2\omega^{[A}\pi^{B]}$; this obeys
\begin{eqnarray}\label{3.14}
\dot J^{5\mu}=-e_2J^{\mu\nu}p_\nu, \qquad
\dot J^{\mu\nu}=-e_2J^{5[\mu}p^{\nu]};
\end{eqnarray}
as well as the position variable
\begin{eqnarray}\label{3.16}
\tilde x^\mu=x^\mu+\frac{1}{2p^2}J^{\mu\nu}p_\nu, \qquad
\dot{\tilde x}^\mu=\tilde ep^\mu.
\end{eqnarray}
We have denoted $\tilde e\equiv e_1+\frac{\hbar e_2}{2mc}$. So the reparametrization-invariant variable $\tilde
x^i(t)$ has a deterministic evolution:
\begin{eqnarray}\label{3.17}
\frac{d\tilde x^i}{dt}=c\frac{\dot{\tilde x}^i}{\dot{\tilde
x}^0}=c\frac{p^i}{p^0}.
\end{eqnarray}
As the classical four-dimensional spin vector, we take the
Pauli-Lubanski vector which has no precession in the free theory
\begin{eqnarray}\label{3.18}
S^\mu=\frac12\epsilon^{\mu\nu\alpha\beta}p_\nu J_{\alpha\beta},
\qquad \dot S^\mu=0
\end{eqnarray}
In the rest frame $p^\mu=(mc, 0, 0, 0)$, it reduces to the three-dimensional rotation generator, $S^0=0$,
$S^i=\frac12mc\epsilon^{ijk}S_{jk}$, as is expected in the non-relativistic limit.

We point out that the second term in Eq. (\ref{3.16}) has the structure typical for non-commutative extensions of the
usual mechanics, see \cite{aad6}.

\section{Conclusion. Pseudo-classical mechanics and the classical limit of the
Dirac equation} Although a true understanding of spin is achieved in the framework of quantum electrodynamics, a lot
of effort has been spent in attempts to construct a mechanical model of a spinning electron, see [1-11, 15-17] and
references therein. The Dirac spinor $\Psi$ can be used to construct the four-dimensional current vector,
$\bar\Psi\gamma^\mu\Psi$, which preserves for solutions to the Dirac equation,
$\partial_\mu(\bar\Psi\gamma^\mu\Psi)=0$. Hence its null-component, $\Psi^\dagger\Psi\ge 0$, admits the probabilistic
interpretation, and we expect that a one-particle sector of the Dirac equation could be described in the framework of
relativistic quantum mechanics (RQM).

However, it is well known that adopting the RQM interpretation, we arrive at a rather strange and controversial
picture [2, 3, 18]. To recall this to mind, we use the Dirac matrices $\alpha^i$ and $\beta$, to represent the Dirac equation
in the form of the Schr\"odinger one
\begin{eqnarray}\label{4.1}
i\hbar\partial_t\Psi=\hat H\Psi, \qquad \hat H= c\alpha^i\hat
p_i+mc^2\beta.
\end{eqnarray}
Then $\hat H$ may be interpreted as the Hamiltonian. If we pass from the Schr\"odinger to the Heisenberg picture, the time
derivative of an operator $a$ is $i\hbar\dot a=[a, H]$. For the basic operators of the Dirac theory we obtain
\begin{eqnarray}\label{4.2}
\dot x_i=c\alpha_i, \qquad i\hbar\dot\alpha_i=2(cp_i-H\alpha_i),
\qquad \dot p_i=0.
\end{eqnarray}
Below we enumerate the inconsistences arising in the RQM interpretation of these equations and show how our pCM allow us to
avoid them.

{\it The wrong balance of the number of degrees of freedom}. Assuming $x$ as the position operator, the first equation
in (\ref{4.2}) implies that the operator $c\alpha^i$ represents the velocity of the particle. Then the physical meaning
of the operator $p^i$ became rather obscure in both the semiclassical and the RQM framework. Using the
quantum-field-theory arguments, Foldy and Wouthuysen [6] argued that the basic operator $x$ which appears in the Dirac
equation does not correspond to the observable quantity. Then they constructed the position operator ${\bf X}^i$ with
reasonable properties.

Our model supports the Foldy-Wouthuysen suggestion. Indeed, we observe that the variable $x^\mu$ is not a
gauge-invariant quantity in our model, so it is not an observable. The observable variable which can be associated with
the particle position is $\tilde x^i(t)$. According to Eq. (\ref{3.17}), $p^i$ determines its velocity. $\tilde
x^\mu(\tau)$ written in Eq. (\ref{3.16}) represents the Lorentz-covariant analog of the operator ${\bf X}^i$ in the
classical theory. We also point out that the Foldy-Wouthuysen transition $x\rightarrow{\bf X}$ corresponds in pCM to
the transition from the gauge-non-invariant to the gauge-invariant variable.

{\it Zitterbewegung}. The equations (\ref{4.2}) can be solved, with the result for $x^i(t)$ being [2, 3]
$x^i=a^i+bp^it+c^i\mbox{exp}(-\frac{2iH}{\hbar}t)$. The last term on the r.h.s. of this equation states that the free
electron experiences rapid oscillations with higher frequency $\frac{2H}{\hbar}\sim\frac{2mc^2}{\hbar}$. It is often
assumed that {\it Zitterbewegung} represents the physically observable motion of a real particle [18]. The analogous
systems that are described by a Dirac-type equation and simulate {\it Zitterbewegung} are under intensive study in
different physical set-ups, including graphene, trapped ions, photonic lattices and ultracold atoms, see \cite{sol} and
references therein.

Our model excludes the {\it Zitterbewegung}, since this represents the dynamics of the unobservable variable $x^\mu$. The
observable variable $\tilde x^i$ moves along a straight line, see Eq. (\ref{3.17}).

{\it Velocity of an electron}. Since the velocity operator
$c\alpha^i$ has eigenvalues $\pm c$, we conclude that a
measurement of a component of the velocity of a free electron is
certain to lead to the result $\pm c$.

In our model, the conjugate momentum $p^\mu$ determines the velocity
of the physical coordinate $\tilde x^\mu$, see Eqs. (\ref{3.16}),
(\ref{3.17}). Then the mass-shell condition (\ref{3.10})
guarantees that the particle cannot exceed the speed of light.

{\it Bargmann-Michel-Telegdi (BMT) equations}. In their seminal work [7], Bargmann, Michel and Telegdi suggested relativistic equations for the classical trajectories and spin precession in uniform fields. The equations practically
exactly reproduced the spin dynamics of polarized beams and agreed with the calculations based on the Dirac theory.
While the BMT model apparently does not imply the Dirac equation, there is a certain relationship between the two schemes.
Namely, the first term of the WKB solution to the Dirac equation can be used to construct the quantities which obey the
BMT equations \cite{rub, raf}.

We have seen above that performing the canonical quantization of our model in the initial variables we arrive at the
Dirac equation. Now we show that physical variables of the model obey the BMT equations.

We take the Hamiltonian of interacting theory in the form
\begin{eqnarray}\label{4.3}
H=\frac{e_1}{2}({\cal P}^2+\frac{e}{2c}F_{\mu\nu}J^{\mu\nu}+m^2c^2)+ \quad \cr \frac{e_2}{2}({\cal P}_\mu
J^{5\mu}+mc\hbar)+ \qquad \qquad \cr
\frac{e_3}{2}(\pi^A\pi_A+a_3)+\frac{e_4}{2}(\omega^A\omega_A+a_4)+\lambda_{ea}\pi_{ea},
\end{eqnarray}
where ${\cal P}_\mu=p_\mu+\frac{e}{c}A_\mu$ is the mechanical momentum. This does not break the local symmetries
presented in the model for the case of uniform electric and magnetic fields. However, the interaction deforms the
physical sector: unambiguous variables of the free theory no longer remain unambiguous in the interacting theory. In
particular, $\tilde x^\mu$, $p^\mu$ and $S^\mu$ now have two-parametric ambiguity. Up to the order $O(\hbar^2)$,
the quantities with one-parametric ambiguity turn out to be
\begin{eqnarray}\label{4.4}
{\bf P}_\mu={\cal P}^\mu-\frac{e}{2c{\cal P}^2}(FJ{\cal P})_\mu, \quad \dot{\bf P}_\mu=-\frac{e}{c}\tilde
eF_{\mu\nu}{\bf P}^\nu; \cr {\bf x}^\mu= x^\mu+\frac{1}{{2\bf P}^2}J^{\mu\nu}{\bf P}_\nu, \qquad ~ ~ \dot{\bf
x}^\mu=\tilde e{\bf P}^\mu; \quad \qquad \cr {\bf S}^\mu=\frac12\epsilon^{\mu\nu\alpha\beta}{\bf P}_\nu
J_{\alpha\beta}, \qquad  \quad ~ ~ \dot{\bf S}^\mu=-\frac{e}{c}\tilde eF^\mu{}_\nu{\bf S}^\nu.
\end{eqnarray}
We have denoted $\tilde e=e_1-\frac{mc\hbar}{2{\cal P}^2}e_2$.
The corresponding reparametrization-invariant variables obey the
BMT equations with $g=2$
\begin{eqnarray}\label{4.5}
\frac{d}{dt}{\bf x}^i=c\frac{{\bf P}^i}{{\bf P}^0}, \quad
\frac{d}{dt}{\bf P}^i=-\frac{e}{{\bf P}^0}F^{i\nu}{\bf P}_\nu, \cr
\frac{d}{dt}{\bf S}^\mu=-\frac{e}{{\bf P}^0}F^{\mu\nu}{\bf S}_\nu.
\qquad
\end{eqnarray}

\section{Acknowledgments}
This work has been supported by the Brazilian foundation FAPEMIG.


\begin{thebibliography}{nn}
\bibitem{1} J. Frenkel, Z. fur Physik {\bf 37}, (1926) 243.
\bibitem{2} E. Schr\"odinger, Sitzunger. Preuss. Akad. Wiss. Phys.-Math. Kl. {\bf 24} (1930) 418.
\bibitem{3} P. A. M. Dirac, {\it The Principles of Quantum Mechanics} (Clarendon Press, Oxford, 1958) p. 261.
\bibitem{4}  M. H. L. Pryce, Proc. Roy. Soc. {\bf A 195} (1948) 62.
\bibitem{5} T. D. Newton and E. P. Wigner, Rev. Mod. Phys. {\bf 21} (1949) 400.
\bibitem{6} L. L. Foldy and S. A. Wouthuysen, Phys. Rev. {\bf 78} (1950) 29.
\bibitem{7} V. Bargmann, L. Michel and V.L. Telegdi, Phys. Rev. Lett. {\bf 2} (1959) 435.
\bibitem{8} A. J. Hanson and T. Regge, Ann. Phys. {\bf 87} (1974) 498.
\bibitem{9} F. A. Berezin and M. S. Marinov, JETP Lett {\bf 21} (1975) 320; Ann. Phys. {\bf 104} (1977) 336.
\bibitem{10} A. Bermudez, M. A. Martin-Delgado, A. Luis, Phys. Rev. {\bf A 77} (2008) 063815.
\bibitem{11} P. Garbaczewski, Phys. Lett. {\bf A 73} (1979) 280.
\bibitem{12} P. A. M. Dirac, Can. J. Math. {\bf 2} (1950) 129; P. A. M. Dirac,
{\it Lectures on Quantum Mechanics} (Yeshiva University, New York, 1964).
\bibitem{13} D. M. Gitman and I. V. Tyutin, {\it Quantization of Fields with
Constraints} (Springer-Verlag, Berlin, 1990) p. 36.
\bibitem{14} A. A. Deriglazov, {\it Classical Mechanics, Hamiltonian and Lagrangian
Formalism} (Springer-Verlag, Berlin Heidelberg, 2010).
\bibitem{15} A. A. Deriglazov, Phys. Lett. {\bf A 376} (2012) 309; arXiv:1106.5228.
\bibitem{16} A. A. Deriglazov, Ann. Phys. {\bf 327} (2012) 398, arXiv:1107.0273.
\bibitem{17} A. A. Deriglazov, B. F. Rizzuti, G. P. Z. Chauca, P. S. Castro, Non-Grassmann mechanical model of the Dirac
equation, arXiv:1202.5757.
\bibitem{18} J. J. Sakurai, {\it Advanced Quantum Mechanics},
(Addison-Wesley Publishing Company Inc., 1967) p. 112, 139.
\bibitem{sol} F. Zahringer, G. Kirchmair, R. Gerritsma, E. Solano, R. Blatt, and C. F. Roos, Nature (London) {\bf 463}, 68 (2010);
\bibitem{rub} S. I. Rubinow and J. B. Keller, Phys. Rev. {\bf 131} (1963) 2789. 
\bibitem{raf} K. Rafanelli and R. Schiller, Phys. Rev. {\bf 135} (1964) B279. 
\bibitem{aad5} A. A. Deriglazov, J. Phys. {\bf A 40} (2007) 11083.
\bibitem{aad6} A. A. Deriglazov, Phys. Lett. {\bf B 530} (2002), 235, Phys. Lett. {\bf B 555} (2003), 83;
J. High Energy Phys. {\bf 3} (2003) 021.

\end{thebibliography}
\end{document}